\documentclass[12pt]{article}
\usepackage[total={6.5in,8.75in}, top=1.2in, bottom=1.2in,left=0.9in, includefoot]{geometry}
\usepackage{graphicx}
\usepackage{amssymb}
\usepackage{axodraw}
\usepackage{color}
\newcommand{\gsim}{\lower.7ex\hbox{$\;\stackrel{\textstyle>}{\sim}\;$}}
\newcommand{\lsim}{\lower.7ex\hbox{$\;\stackrel{\textstyle<}{\sim}\;$}}
\def\hnu{h_{\nu}}
\def\hp{h_{\rm p}}

\title{The Minimal Phantom Sector of the Standard Model:\\
Higgs Phenomenology and Dirac Leptogenesis}
\author{D.~G.~Cerde\~no, A.~Dedes and T.~E.~J. Underwood\\[4mm]
{\it \small Institute for Particle Physics
Phenomenology (IPPP),}\\{\it \small University of  Durham, DH1 3LE, UK}}
\date{14$^{\rm th}$ July 2006}        

\begin{document}

\begin{flushright}
IPPP/06/46\\
DCPT/06/92\\
July 2006
\end{flushright}

\bigskip

\begin{center}
{\LARGE The Minimal Phantom Sector of the Standard Model:\\[4mm]
Higgs Phenomenology and Dirac Leptogenesis}\\[6mm]

{\large D.~G.~Cerde\~no, A.~Dedes and T.~E.~J. Underwood\\[3mm]}
{\it \small Institute for Particle Physics Phenomenology (IPPP),}\\
{\it \small University of  Durham, DH1 3LE, UK.}\\[10mm]
\end{center}


\begin{abstract}
  We propose the minimal, lepton-number conserving, SU(3)$_{\rm
    c}\times$SU(2)$_L\times$U(1)$_Y$ gauge-singlet, or phantom, extension of
  the Standard Model.
  The extension is natural in the sense that all couplings are of ${\cal
    O}$(1) or forbidden due to a phantom sector global $U(1)_{\rm D}$
  symmetry, and basically imitates the standard Majorana see-saw mechanism.
  Spontaneous breaking of the $U(1)_{\rm D}$ symmetry triggers consistent
  electroweak gauge symmetry breaking only if it occurs at a scale compatible
  with small Dirac neutrino masses and baryogenesis through Dirac
  leptogenesis.
  Dirac leptogenesis proceeds through the usual out-of-equilibrium decay
  scenario, leading to left and right-handed neutrino asymmetries that do not fully
  equilibrate after they are produced.
%
  The model contains two physical Higgs bosons and a massless Goldstone boson.
  The existence of the Goldstone boson suppresses the Higgs to $bb$ branching
  ratio and instead the Higgs bosons will mainly decay to invisible Goldstone
  and/or to visible vector boson pairs. In a representative scenario, we
  estimate that with $30$~fb$^{-1}$ integrated luminosity, the LHC could
  discover this invisibly decaying Higgs, with mass $\sim$120 GeV.  At the
  same time a significantly heavier, partner Higgs boson with mass $\sim$210
  GeV could be found through its vector boson decays. Electroweak constraints
  as well as astrophysical and cosmological implications are analysed and
  discussed.
\end{abstract}


\subsection{\it Introduction}

The Standard Model (SM) has just two openings where renormalisable operators can
be added which couple SU(3)$_{\rm c}\times$SU(2)$_L\times$U(1)$_Y$ gauge
singlet fields to SM fields. One place is the super-renormalisable Higgs mass
term \cite{wilczek}, the other place is the lepton-Higgs Yukawa interaction
$\bar{L}\widetilde{H}$. What would happen if we filled in
these gaps?

There is no physical evidence, as yet, to suggest that $B-L$ (baryon -- lepton
number) is not a good symmetry of nature. The SM preserves $B-L$, so we will
choose to extend the SM in a $B-L$ preserving way. However, overwhelming
evidence supporting small, non-zero neutrino masses does exist \cite{numass}.
We are therefore led to build a model with Dirac masses for the neutrinos and
see if it is possible to create the observed baryon asymmetry of the Universe
within this set up. Ideally, we should also strive to build a natural model,
both in the 't~Hooft sense and the aesthetic sense. In particular, Yukawa
couplings should be either ${\cal O}$(1) or strictly forbidden.

Following this approach, we augment the SM with two SU(3)$_{\rm
  c}\times$SU(2)$_L\times$U(1)$_Y$ gauge singlet fields, a complex scalar
$\Phi$ and a Weyl fermion $s_R$. These fields will provide the link between
the SM and a phantom gauge singlet sector.
\begin{equation}
-{\cal L}_{\rm link} = \Big(\,\hnu\,\overline{l_{L}} \cdot
\tilde{H}\,s_{R} \ +\ {\rm H.c.}\,\Big)
\ -\ \eta\,H^\dagger H\,\,\Phi^*\,\Phi\,,
\label{Llink}
\end{equation}
where $\hnu$ and $\eta$ are dimensionless couplings of ${\cal O}$(1) in
line with our naturalness criterion. The field $H$ (or $\tilde{H}=i
\sigma_2 H^*$) is the standard $SU(2)_L$ Higgs doublet\footnote{In our
  notation is $H^T=(H^{+}, H^{0})$ and $\tilde{H}^T=(H^{0*},
  -H^{-})$.} responsible for spontaneous electroweak, $SU(2)_L\times U(1)_Y$, gauge 
  symmetry breaking. Note that
$s_R$ must carry lepton number $L=1$.

In this form, the model is incomplete because neutrinos would have large,
electroweak-scale masses. However, we have so far ignored the purely phantom,
gauge singlet sector of the model. Here we add a phantom right-handed neutrino
$\nu_R$, and $s_L$ the partner of the Weyl field $s_R$.  These fields will
also carry lepton number. The fermionic part of the phantom sector therefore
contains
\begin{equation}
-{\cal L}_{\rm p} = \hp\,\Phi \,\overline{s_{L}}\,\nu_{R} 
\ + \ M \,\overline{s_{L}} \, s_{R}  \ + \ {\rm H.c.}
\label{Lphant}
\end{equation}
where $\hp$ is a general complex Yukawa coupling of ${\cal O}$(1). Other
possible lepton number conserving terms,
\begin{equation}
\overline{l_L} \: \tilde{H} \:  \nu_R \ + \ M' \:  \overline{s_L} \: \nu_R \ + \ 
\Phi \: \overline{s_L} \: s_R \ + \ {\rm H.c.} \ +\ \ldots \;,
\label{noterms}
\end{equation}
are forbidden when imposing a global $U(1)_{\rm D}$ symmetry, under which only
the fields,
\begin{eqnarray}
\nu_R \to e^{i\alpha} \nu_R \qquad , \qquad \Phi \to e^{-i\alpha} \Phi \;,
\label{global}
\end{eqnarray}
transform non-trivially. This choice for the phantom sector is purely motived
by the need for the simplest model leading to small, Dirac neutrino masses.  A
crucial point to notice here is that {\it the spontaneous symmetry breaking of
  the global $U(1)_{\rm D}$ will trigger consistent electroweak gauge symmetry
  breaking through the last term in ${\cal L}_{\rm link}$ provided that
  $\langle \Phi \rangle \equiv \sigma \sim v$, with $v$ being the vacuum
  expectation value (vev) of $H$}.

The model 
\begin{equation}
{\cal L} \ =\ {\cal L}_{\rm SM} \ +\ {\cal L}_{\rm link} \ +\ {\cal L}_{\rm p}\,,
\label{model}
\end{equation}
can be trivially embedded into Grand Unified Theories (GUTs) and can be
supersymmetrized.  Part of the model was first presented in the literature by
Roncadelli and Wyler~\cite{RW}, who were motivated by the need for a model
with naturally small Dirac neutrino masses~\cite{Ma}. Notice that its structure is different (and
much simpler) than the ones exploited recently by \cite{Murayama,Abel,Boz}, though the
latter are supersymmetric. 

We will confine our discussion to a three generation neutrino model. We will
therefore add three generations of the $s_R$, $s_L$ and $\nu_R$. For
simplicity we will consider just one copy of the complex scalar $\Phi$.  This
is the {\it Minimal Phantom Sector of the SM} consistent with naturally small
Dirac neutrino masses, and as we shall see shortly, provides an explanation
for the baryon asymmetry of the Universe and has Higgs phenomenology
strikingly different to that of the SM.

\subsection{\it  Neutrino Masses}

In this section we briefly repeat the main points of ~\cite{RW}. Notice that
we are free to work in the basis where the Dirac mass matrix $M$ in
(\ref{Lphant}) is diagonal.
%
After spontaneous $U(1)_{\rm D}$ symmetry breaking, the Lagrangian (\ref{model}) results in
Dirac-neutrino effective mass terms of the form, 
$\overline{{\bf \nu_L'}} \: {\bf m_\nu} \: {\bf \nu_R'} \ + \  \overline{s_L'} \:
 {\bf m_N \: s_R'}$, where up to terms ${\cal O}({\bf M^{-2}}$), we obtain
\begin{eqnarray}
{\bf m_\nu} \ = - \ {\bf m}\:  {\bf \hat{M}^{-1} }\:  {\bf m_{\rm p}} \qquad , \qquad  
 {\bf m_N} = {\bf \hat{M}} \;,
 \label{numass}
 \end{eqnarray}
 with ${\bf m} = {\bf \hnu} v$ and ${\bf m_p} = {\bf \hp} \sigma$ being
 $3\times 3$ matrices and with neutrino mass eigenstates
\begin{eqnarray}
{\bf  \nu_L'} \  &=& \  {\bf \nu_L} \ - \ {\bf m \: \hat{M}^{-1} \: s_L} \qquad ,  \qquad 
{\bf s_L'} \ = \  {\bf  \hat{M}^{-1} \: m^\dagger  \:  \nu_L} \ + \ {\bf s_L} \nonumber  \\[2mm]  
{\bf \nu_R'} \  &=& {\bf \nu_R} \ - \  {\bf m_p^\dagger \: \hat{M}^{-1} \: s_R} \qquad , \qquad
{\bf s_R'} \ = \ {\bf \hat{M}^{-1}\:  m_p \: \nu_R} \ + \  {\bf s_R} \; .
\label{nueigen}
\end{eqnarray}
Bold face letters denote $3\times 3$ matrices and column vectors. From
(\ref{numass}) we obtain a typical seesaw spectrum with light and heavy Dirac
neutrinos with masses ${\bf m_\nu}$ and ${\bf \hat{M}}$, respectively.

\begin{figure}
\begin{center}
  \begin{picture}(130,53) (30,-24)
    \SetWidth{0.8}
    \SetColor{Black}
    \ArrowLine(60,8)(90,8)
    \ArrowLine(30,8)(60,8)
    \ArrowLine(120,8)(150,8)
    \Line(88,6)(92,10)\Line(88,10)(92,6)
    \DashLine(60,8)(60,-22){4}
    \ArrowLine(90,8)(120,8)
    \Line(58,-24)(62,-20)\Line(58,-20)(62,-24)
    \DashLine(120,8)(120,-22){4}
    \Line(118,-24)(122,-20)\Line(118,-20)(122,-24)
    \Text(66,-17)[lb]{\normalsize{\Black{$\Phi$}}}
    \Text(126,-17)[lb]{\normalsize{\Black{$H$}}}
    \Text(130,13)[lb]{\normalsize{\Black{$\nu_L$}}}
    \Text(100,13)[lb]{\normalsize{\Black{$s_R$}}}
    \Text(70,13)[lb]{\normalsize{\Black{$s_L$}}}
    \Text(40,13)[lb]{\normalsize{\Black{$\nu_R$}}}
  \end{picture}
\end{center}
\caption{\em The diagram responsible for light Dirac neutrino masses in the model (\ref{model}).}
\label{fig:feynman}
\end{figure}
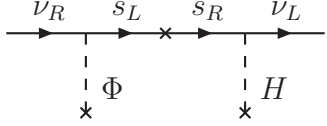

The physical neutrino masses will then result from the final rotation ${\bf
  \hat{m}_\nu }= {\bf A^\dagger { m_\nu} B}$ where ${\bf A, B}$ are unitary
matrices. In a basis where the charged lepton Yukawa couplings, ${\bf h}_e$
are diagonal, the matrix ${\bf A}$ will just be the usual PMNS
matrix~\cite{MNS}, measured by neutrino flavour oscillation experiments.
Minimal mass matrices ${\bf m_\nu}$ for Dirac neutrinos have been recently
classified in~\cite{HR} and can be exploited to shed light onto the connection
between CP-violation and Dirac leptogenesis that follows in the next section.
The reader has possibly already realized that the model in (\ref{model}) is a
simple realization of the Froggatt-Nielsen~\cite{FN} mechanism, usually invoked
to generate large hierarchies in quark masses. In this model neutrino masses
are generated by the Feynman diagram in Fig.~\ref{fig:feynman}
and are given by
\begin{eqnarray}
M_{ij} = g_{ij} \: \epsilon^{a_i + b_j} \;,
\end{eqnarray}
with $\epsilon = \sigma/M$ and $a_i + b_j=1$, corresponding to the difference
between the $U(1)_{\rm D}$-charges of the left and the right handed neutrino.
The matrix $g_{ij}$ is a general matrix, in our case a product of the matrices
${\bf \hnu} v$ and ${\bf \hp}$. In contrast with the quark case where
$\epsilon$ naturally explains the large hierarchy in quark masses, here the
smallness of $\epsilon$ explains the relative smallness of neutrino masses.
Assuming that the Yukawa couplings are perturbative and of order one,
with $v=175$ GeV and $m_\nu = 0.05$ eV we find the ratio $\epsilon = \sigma /M$, needs to be
\begin{eqnarray}
 \epsilon \ \simeq \ 3 \times 10^{-13} \;.
 \end{eqnarray}
 In the next section we will see that it is possible to achieve this in a
 model consistent with our naturalness criterion and successful baryogenesis
 \cite{Lindner}. Significantly, in this model there is no neutrinoless double
 beta decay, however, we will later examine the consequences of the phantom
 sector for Higgs searches at the LHC.



\subsection{\it Baryogenesis}
\subsubsection{\it  Dirac Leptogenesis }

Although $B-L$ is preserved exactly in this model, we will see that
baryogenesis through (Dirac) leptogenesis is still possible \cite{Lindner}.
Just as in the SM \cite{thooft}, in this model the combination $B+L$ is
anomalous and at low temperatures $B+L$ violation proceeds through tunnelling
and is un-observably small. At higher temperatures, close to and above the
critical temperature for electroweak symmetry breaking, $T_c
\stackrel{>}{_\sim} 150$~GeV, thermal fluctuations allow field configurations
to pass over the `sphaleron barrier', leading to rapid $B+L$ violation
\cite{KRS}.

It is important to note that the rapid $B+L$ violating processes do
not directly affect right-handed gauge singlet particles. Large Yukawa
couplings between the SM quarks and charged leptons will however tend
to equilibrate asymmetries in the left and right sectors of the model,
depleting an overall `right-handed' $B+L$ as an overall `left-handed'
$B+L$ is depleted via `sphaleron effects'.

The crucial idea behind Dirac leptogenesis (or Dirac neutrinogenesis) is that
the small {\em effective} Yukawa couplings between the SM Higgs and the left and right handed
neutrinos could prevent asymmetries in the neutrino sector of the model from
equilibrating. Therefore, even in a model conserving total lepton number, a
left-handed $B-L$ asymmetry could be produced at the same time as an opposite,
right-handed $B-L_{\nu_R}$ asymmetry. The left-handed $B-L$ asymmetry would
then lead to an overall baryon asymmetry just as in Majorana leptogenesis.
Clearly, for this to work the {\em effective} neutrino Yukawa couplings must be small enough
to keep the left and right lepton asymmetries from equilibrating until (at
least) after the electroweak phase transition, when the sphaleron processes
linking the baryon and lepton asymmetries would have dropped out of thermal
equilibrium.  This mechanism even works when the initial, overall $B = L = 0$.
It is especially interesting to note that this mechanism links the baryon
asymmetry directly to the smallness of the Dirac neutrino masses.

At temperatures above $T_c$, when sphaleron and other SM processes can
maintain most SM species in thermal equilibrium it is possible to derive
relations amongst the chemical potentials of the various particle species
\cite{HT}.  Since we demand the right handed neutrinos be out of thermal
equilibrium we can ignore their contribution for the moment, leading to the
usual relation between baryon and lepton number used in Majorana leptogenesis
\begin{equation}
\label{convBLrel}
Y_{B} = \frac{28}{79}\,(Y_{B} - Y_{L_{\rm SM}})\,,
\end{equation}
where $Y_{X}$ refers to the `asymmetry in X' to entropy ratio and $L_{\rm SM}$
refers to the lepton number held in SM particles (not including the right
handed neutrinos). If we now consider the case where, initially, the total
$B-L$ was zero, we have the relation
\begin{equation}
Y_{B} - Y_{L_{\rm SM}} - Y_{L_{\nu_R}} = 0,
\end{equation}
which in conjunction with equation (\ref{convBLrel}) leads to
\begin{equation}
Y_{B} = -\frac{28}{79}\,Y_{L_{\nu_R}},
\end{equation}
showing that just over a quarter of the right handed neutrino asymmetry is
converted into a baryon asymmetry.

\subsubsection{\it CP-violation}

As is well known, in order to generate a particle-antiparticle
asymmetry in the early universe, the three Sakharov criteria must be
fulfilled \cite{Sak}. Particularly relevant to this model are the
requirements for a departure from thermal equilibrium and
CP-violation.

\begin{figure}
\begin{center}
\begin{picture}(273,85) (17,-18)
    \SetWidth{0.5}
    \SetColor{Black}
    \Text(94,5)[lb]{\normalsize{{$\Phi$}}}
    \Text(95,47)[lb]{\normalsize{{$\nu_{R\,k}$}}}
    \Text(17,26)[lb]{\normalsize{{$S_i$}}}
    \Text(122,26)[lb]{\normalsize{{$S_i$}}}
    \SetWidth{0.8}
    \ArrowLine(135,30)(165,30)
    \ArrowArcn(180,30)(15,180,-0)
    \DashArrowArc(180,30)(15,-180,0){4}
    \ArrowLine(195,30)(225,30)
    \ArrowLine(225,30)(255,50)
    \DashArrowLine(225,30)(255,10){4}
    \ArrowLine(30,30)(60,30)
    \ArrowLine(60,30)(90,50)
    \DashArrowLine(60,30)(90,10){4}
    \Text(260,47)[lb]{\normalsize{{$\nu_{R\,k}$}}}
    \Text(259,5)[lb]{\normalsize{{$\Phi$}}}
    \Text(205,35)[lb]{\normalsize{{$S_j$}}}
    \Text(175,51)[lb]{\normalsize{{$L_l$}}}
    \Text(177,3)[lb]{\normalsize{{$H$}}}
    \Text(51,-18)[lb]{\normalsize{{$(a)$}}}
    \Text(185,-18)[lb]{\normalsize{{$(b)$}}}
  \end{picture}
\end{center}
\caption{\it Feynman diagrams for the decay of the heavy gauge singlet
  $S_i$ into a $\nu_{R\,k}$ and a $\Phi$. The CP-asymmetry in this decay is
  due to the interference between the tree-level diagram $(a)$ and the 1-loop
  self-energy diagram $(b)$.}
\label{fig:CP}
\end{figure}
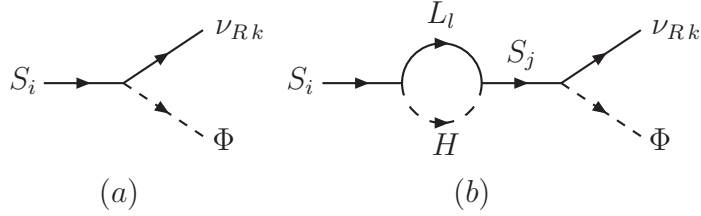

In analogy with conventional leptogenesis, in this model, CP-violation and a
departure from thermal equilibrium could arise during the decays of the heavy
Dirac $S \equiv s_L + s_R$ particles. In particular, CP-violation would originate
through the interference between a tree level decay and a 1-loop self-energy
diagram as shown in Fig.~\ref{fig:CP}.
If we define a `right' CP-asymmetry as
\begin{equation}
\delta_{R\,i} = \frac{\sum_k \Big(\Gamma(S_i \to \nu_{R\,k}\,\Phi) -
\Gamma(\bar{S}_i \to \bar{\nu}_{R\,k}\,\Phi^*)\Big)}{\sum_j \Gamma(S_i \to
  \nu_{R\,j}\,\Phi) + \sum_l \Gamma(S_i \to L_l\,H)}\,,
\label{defdeltaR}
\end{equation}
where $\Gamma(S_i \to \nu_{R\,k}\,\Phi)$ is the rate of the decay process
$S_i \to \nu_{R\,k}\,\Phi$ etc. $L_l$ and $H$ represent lepton and SM
Higgs SU(2)$_L$ doublets respectively. In addition, unitarity and CPT
conservation provide the following useful relation
\begin{equation}
\sum_j \Gamma(S_i \to \nu_{R\,j}\,\Phi) + \sum_l \Gamma(S_i \to
L_l\,H)
=
\sum_{j'} \Gamma(\bar{S}_{i} \to \bar{\nu}_{R\,j'}\,\Phi^*) + \sum_{l'} \Gamma(\bar{S}_i \to
\bar{L}_{l'}\,H^\dagger)\,,
\end{equation}
which leads to the relation $\delta_{L\,i} = - \delta_{R\,i}$, where
$\delta_{L\,i}$ is defined in analogy with equation (\ref{defdeltaR}).

We also define right and left branching ratios as
\begin{equation}
B_{R\,i} = \frac{\sum_k \Big(\Gamma(S_i \to \nu_{R\,k}\,\Phi) +
\Gamma(\bar{S}_i \to \bar{\nu}_{R\,k}\,\Phi^*)\Big)}{\sum_j \Gamma(S_i \to
  \nu_{R\,j}\,\Phi) + \sum_l \Gamma(S_i \to L_l\,H)}\,,
\end{equation}
and $B_{L\,i} = 2 - B_{R\,i}$.

In the limit that the $M_i$ are hierarchical, the lepton asymmetry is
generated via the decay of the lightest $S$ and $\bar{S}$. In this case, the
CP-asymmetry (\ref{defdeltaR}) is given by
\begin{equation}
\delta_{R1} \simeq \frac{1}{8\pi}\,\sum_{j}\,\frac{M_{1}}{M_{j}}\,\frac{{\rm
    Im}\Big[({\bf \hp\,\hp^{\dagger}})_{1j}\,({\bf
    \hnu^{\dagger}\,\hnu})_{j1}\Big]}{({\bf \hp\,\hp^{\dagger}})_{11} + ({\bf
    \hnu^{\dagger} \hnu})_{11}}\,.
\label{CPasym}
\end{equation}
It should be noted that the structure of the amplitudes leading to
CP-violation in this scenario are similar to the self-energy contribution to
the CP-asymmetry in Majorana leptogenesis.  Therefore, in the limit of
quasi-degenerate masses for the $S_i$, a resonant enhancement of $\delta_R$
should be possible \cite{Liu,Apostolos}. We will however leave this case to be
considered elsewhere.
Using equation (\ref{numass}), (\ref{CPasym}) can be rewritten
\begin{equation}
\delta_{R1} \simeq -\frac{1}{8\pi}\,\frac{M_{1}}{v\,\sigma}\,\frac{{\rm
    Im}\Big[({\bf \hp\,m_\nu^\dagger\,\hnu})_{11}\Big]}{({\bf \hp\,\hp^{\dagger}})_{11}
    + ({\bf \hnu^{\dagger} \hnu})_{11}}\,.
\end{equation}
Following the approach of \cite{CI}, the most general ${\bf \hnu}$ and ${\bf
  \hp}$ can be parameterised in the following way
\begin{eqnarray}
{\bf \hnu} \!& = &\! \frac{1}{v}\,{\bf A\,D_{\sqrt{\hat{m}_\nu}}\,W\,
D_{\sqrt{\hat{M}}}}\,,\\
{\bf \hp} \!& = &\! \frac{1}{\sigma}\,{\bf D_{\sqrt{\hat{M}}}\,X^\dagger\,D_{\sqrt{\hat{m}_\nu}}\,B^\dagger}\,,
\end{eqnarray}
where $\bf W$ and $\bf X$ are general $3\times3$ matrices satisfying the
condition $\bf W\,X^\dagger = 1$ and $\bf D_{\sqrt{Z}} = +\sqrt{Z}$ for the
diagonal matrix $\bf Z$. This allows $\delta_{R1}$ to be written as
\begin{equation}
\delta_{R1} \simeq -\frac{1}{8\pi}\,\frac{M_{1}}{v\,\sigma}\,\frac{{\rm Im}\Big[({\bf
  X^\dagger\,\hat{m}_\nu^2\,W})_{11}\Big]}{({\bf X^\dagger\,\hat{m}_\nu\,X})_{11}+({\bf W^\dagger\,\hat{m}_\nu\,W})_{11}}\,.
\end{equation}
It is now straightforward to show, in analogy with \cite{DI}, that
$|\delta_{R1}|$ is bounded from above by
\begin{equation}
|\delta_{R1}| \stackrel{<}{_\sim}
 \frac{1}{16\pi}\,\frac{M_{1}}{v\,\sigma}\,(m_{\nu_3} - m_{\nu_1})\,.
\label{CPbound}
\end{equation}
Again, we should stress that this bound could be violated grossly when the
$M_i$ are nearly degenerate, and that this caveat should be noted when this
bound is used later. As we seek the most minimal model, without additional
flavour symmetries in the phantom sector, we will not consider this resonant
scenario here.

\subsubsection{\it  Out of equilibrium decays}

In the early Universe, if the expansion rate is faster than the
interaction rate of a particle species then this species can become
decoupled from the thermal bath. This statement can be quantified by
considering the ratio of the particle interaction rate $\Gamma(T)$, to
the Hubble expansion rate $H(T)$. If this ratio is less than 1, then
the species evolves out of thermal equilibrium \cite{KT}.

If we require successful Dirac leptogenesis, the distributions of left and
right handed neutrinos should be prevented from coming into thermal
equilibrium from the moment a $B-L_{\nu_R}$ asymmetry is created until after
the electroweak phase transition, when the rate for $B+L$ violating processes
will be much smaller than the expansion rate of the universe.

Processes leading to left-right equilibration include
$LH\leftrightarrow\Phi\,\nu_R$ mediated by the $s$-channel exchange of an $S_i$,
$L\,\bar{\nu}_R\leftrightarrow H\,\Phi$ and $L\Phi\leftrightarrow\nu_R\,H$ mediated
by the $t$-channel exchange of an $S_i$. Approximately, at high temperatures
these processes have a rate
\begin{equation}
\Gamma_{L\leftrightarrow R}\,(T) \sim \frac{|\hnu|^2\,|\hp|^2}{M_1^4}\,T^5\,,
\end{equation}
which should be compared to the Hubble parameter in the relevant
radiation dominated era
\begin{equation}
H(T) = \sqrt{\frac{8 \pi^3 g_*}{90}}\,\frac{T^2}{M_P}\,,
\end{equation}
where $g_* \simeq 114$ is the effective number of relativistic degrees of
freedom in the SM plus 3 $\nu_R$ and 1 complex $\Phi$, and $M_P =
1.2\times10^{19}$~GeV is the Planck mass. The strongest constraint will come
from the highest temperatures when $T \simeq M_1$, i.e. those at which the
asymmetry is generated;
\begin{equation}
\frac{|\hnu|^2\,|\hp|^2}{M_1} \lsim \frac{1}{M_P}\,\sqrt{\frac{8 \pi^3 g_*}{90}}\,.
\end{equation}
To more accurately consider this constraint we need to solve the appropriate
Boltzmann equations (discussed next). The dominant contribution to left-right
equilibration will come from the inverse decay and subsequent decay of a real
$S_1$ or $\bar{S_1}$.

In this model, a $L_{\nu_R}$ asymmetry can be generated via the standard
out-of-equilibrium decay scenario, in analogy with various GUT baryogenesis
scenarios and Majorana leptogenesis \cite{GUT,Maj}. In the following, we will
consider the asymmetry to be generated solely during the decays of the
lightest $S$.  Pre-existing asymmetries from, for example, the decays of the
heavier $S$s will be treated as initial conditions.

We will assume at $T\gsim M_1$ the abundance of $S_1$ and $\bar{S_1}$ is
thermal (we will relax this assumption later), and the number density of
$S_1$, $n_{S_1} \simeq n_{\bar{S}_1} \simeq n_\gamma$, where $n_\gamma$ is the
number density of photons. As the Universe expands and cools the number
density of the $S_1$ ($\bar{S_1}$) must rapidly decrease if they are to remain
in thermal equilibrium below $T \simeq M_1$. If the interactions allowing this
(primarily $S_1$ ($\bar{S_1}$) decays) are slow compared to the expansion rate
of the Universe then the $S_1$ and $\bar{S}_1$ abundances will depart from
their thermal equilibrium values. When the $S_1$ ($\bar{S}_1$) eventually
decay, the rates of back-reactions such as inverse decays will be suppressed
by the relatively low temperature $T\ll M_1$ and the resulting $L_{\nu_R}$
asymmetry will be \cite{KT}
\begin{eqnarray}
Y_{L_{\nu_R}} \equiv \frac{n_{L_{\nu_R}}}{s} \ \simeq\
\frac{\delta_R\,n_{S_1}}{g_*\,n_\gamma}\ \simeq\ \frac{\delta_R}{g_*}\,.
\label{simple}
\end{eqnarray}
We can define a parameter $K$ such that
\begin{equation}
K \equiv \frac{\Gamma(S_1\to \nu_R \Phi) + \Gamma(S_1\to L H)}{H(T=M_{1})} =
\Big[({\bf \hp \hp^\dagger})_{11} + ({\bf \hnu^\dagger
\hnu})_{11}\Big]\,\frac{M_P}{16\pi\,M_{1}}\, \sqrt{\frac{90}{8 \pi^3 g_*}}\,,
\end{equation}
where $K \ll 1$ signifies that $S_1$ is completely out of thermal equilibrium
at $T=M_{1}$; the `drift and decay' limit. We can re-cast this constraint in
terms of `effective neutrino masses' \cite{BP} to make the connection with
light neutrino data more transparent. Defining the effective neutrino mass as
\begin{equation}
\widetilde{m} \ \equiv\ \Big[({\bf \hp \hp^\dagger})_{11} + ({\bf \hnu^\dagger
\hnu})_{11}\Big]\,\frac{v\,\sigma}{M_1} \ =\ K\,v\,\sigma\,\frac{16\pi}{M_P}\,
\sqrt{\frac{8\pi^3\,g_*}{90}}\,,
\end{equation}
we see that $K<1$ is satisfied for $\widetilde{m}<m_*$ where
\begin{equation}
m_* \ =\ v\,\sigma\,\frac{16\pi}{M_P}\,
\sqrt{\frac{8\pi^3\,g_*}{90}}\,.
\end{equation}
The connection between $\widetilde{m}$ and the physical light neutrino masses
is clearly model dependent, and most applicable when $({\bf \hp
  \hp^\dagger})_{11} \simeq ({\bf \hnu^\dagger \hnu})_{11}$.

Finally, we can also introduce an efficiency parameter $\kappa$, such that the
$L_{\nu_R}$ yield is given by
\begin{equation}
Y_{L_{\nu_R}} = \frac{\delta_{R1}\,\kappa}{g_*}\,.
\label{kappa}
\end{equation}
For an initially thermal population of $S_1$ and when $K \ll 1$ the
efficiency $\kappa \simeq 1$.

If $K > 1$, Dirac leptogenesis can still be successful if the CP-asymmetry
$\delta_R$ is large enough. However, the simple estimate of the lepton
asymmetry, equation (\ref{simple}) will no longer be valid, and the Boltzmann
equations (BEs) should be solved. There are 4 coupled Boltzmann equations
relevant to this scenario, two for the $S_1$ total abundance and asymmetry,
one for the asymmetry in $L_L$ and one for the asymmetry in $L_{\nu_R}$.
Lepton number conservation means that only three of the asymmetry BEs are
independent, we choose to eliminate the equation for $L_{\nu_R}$. The
derivation of the BEs has been extensively covered in the literature, see for
example \cite{BP,KolbWolf,GNRRS}, so we will just write down the set of
simplified equations for this scenario. The BEs read
\begin{eqnarray}
\frac{d \eta_{\Sigma S_1}}{dz} \!& = &\! \frac{z}{H(z=1)}\Bigg[
2 \ -\ \frac{\eta_{\Sigma S_1}}{\eta_{S_1}^{\rm eq}} \ +\
\delta_R\,\Bigg(\,\frac{3\,\eta_{\Delta L}}{2} \ +\ \eta_{\Delta S_1}\,\Bigg)\Bigg]
\,\Gamma^{D1}\,,\nonumber\\
\frac{d \eta_{\Delta S_1}}{dz} \!& = &\! \frac{z}{H(z=1)}\Bigg[
\eta_{\Delta L} \ -\ \frac{\eta_{\Delta S_1}}{\eta_{S_1}^{\rm eq}}
\ -\ B_R\,\Bigg(\,\frac{3\,\eta_{\Delta L}}{2} \ +\ \eta_{\Delta S_1}\,\Bigg)\Bigg]
\,\Gamma^{D1}\,,\nonumber\\
\frac{d \eta_{\Delta L}}{dz} \!& = &\! \frac{z}{H(z=1)}\Bigg\{\,\Bigg[
\delta_R\,\Bigg(\,1\ -\ \frac{\eta_{\Sigma S_1}}{2 \eta_{S_1}^{\rm eq}}\,\Bigg)
\ -\ \Bigg(\,1-\frac{B_R}{2}\,\Bigg)\,\Bigg(\,\eta_{\Delta L}\ -\ \frac{\eta_{\Delta S_1}}{\eta_{S_1}^{\rm eq}}\,\Bigg)\Bigg]
\,\Gamma^{D1}\nonumber\\
\!& &\! \qquad\qquad\qquad -\ \Bigg(\,\frac{3\,\eta_{\Delta L}}{2} \ +\ \eta_{\Delta
  S_1}\,\Bigg)\,\Gamma^W\,\Bigg\}
\label{BEs}
\end{eqnarray} 
where $\eta_{\Sigma S} = (n_{S} + n_{\bar{S}})/n_\gamma$, $\eta_{\Delta S}
= (n_{S} - n_{\bar{S}})/n_\gamma$, $z = M_1/T$ and
\begin{eqnarray}
\Gamma^{D1} \!& = &\! \frac{1}{n_\gamma}\,\bigg[\Gamma(S_1\to \nu_R\Phi)+\Gamma(S_1\to LH)\bigg]\,g_{S_1}\,
\int \!\frac{d^3 {\bf p}}{(2\pi)^3}\,\frac{M_1}{E_{S_1}}\,
e^{-E_{S_1}/T}\,,\nonumber\\
\!& = &\! \bigg[\Gamma(S_1\to \nu_R\Phi)+\Gamma(S_1\to L H)\bigg]\ 
\frac{z^2}{2} K_1 (z)\,,
\end{eqnarray}
where $g_{S_1} = 2$ is the number of internal degrees of freedom of $S_1$,
$E_{S_1} = \sqrt{{\bf p}^2 + M_1^2}$ and $K_1 (z)$ is a 1st order modified
Bessel function. Decays and inverse decays of $S_1$ and $\bar{S_1}$ are
included through terms proportional to $\Gamma^{D1}$. Notice that these terms
also include the most important CP-violating $2\leftrightarrow 2$ scattering
contribution coming from the subtraction of real intermediate states from the
process $LH \leftrightarrow \nu_R \Phi$ \cite{KolbWolf}. This subtraction is
necessary to ensure unitarity and CPT are respected by avoiding double
counting of processes in the BEs.

$\Gamma^W$ parameterises the `wash-out' due to processes of higher order in
the Yukawa couplings which will tend to equilibrate the $L$ and $\nu_R$
asymmetries, after the above subtraction of possible real intermediate states
has been carried out. These process will be predominantly mediated by the
off-shell exchange of $S_{1,2,3}$ and will therefore be highly model dependent
\cite{BP,GNRRS}. For small Yukawa couplings where $|\hp|^2 \simeq |\hnu^2| <
1$, the processes $L\,\bar{\nu}_R\leftrightarrow H\,\Phi$ and
$L\Phi\leftrightarrow\nu_R\,H$ will be negligible compared to decays and
inverse decays. The contribution from the off-shell process $LH
\leftrightarrow \nu_R \Phi$ is bounded from above by $\Gamma^{D1}$ for the
region around $z\sim1$, if the $S_i$ have a reasonably large hierarchy in mass.
We will therefore make the conservative approximation that $\Gamma^W =
\Gamma^{D1}$.

Other $2\leftrightarrow 2$ processes, for example those involving an $S_1$ in
the initial or final state, have been neglected since their main contribution
is at $T\gsim M_1$ where they would act to help create an initially thermal
population of $S_1$ \cite{BP}. These processes would tend to increase the
leptogenesis efficiency in scenarios with zero initial abundance of $S_1$,
when $K\ll1$. The bounds derived later will depend on the leptogenesis
efficiency at large values of $K\gg 1$, therefore we expect these processes to
have a negligible impact. The same point can be made with regard to thermal
effects, which would kinematically block the decays of $S_1$ at temperatures
$T\gsim M_1$ \cite{GNRRS}. As we consider scenarios in the large $K$ regime,
the baryon asymmetry is predominantly determined by processes at $T\lsim M_1$,
leading to only small finite temperature corrections to our $T=0$ estimates.

Clearly, our treatment of the BEs is approximate. We have also neglected
lepton flavour effects, coming from the charged lepton Yukawa couplings
\cite{SDR} and the neutrino Yukawa couplings \cite{APTU}. These effects could
be large for scenarios with $M_1<10^{12}$~GeV, when the charged lepton Yukawa
couplings are in thermal equilibrium, or when there is only a mild hierarchy
in the $M_i$s. These effects could lead to differences in the final baryon
asymmetry of up to an order of magnitude, however they are highly model
dependent and would require a study beyond the scope of this paper. Since the
purpose of our discussion here is to provide an existence proof and a
reasonable estimate of the baryon asymmetry in very general scenarios our
approach is expected to be accurate enough.

\begin{figure}
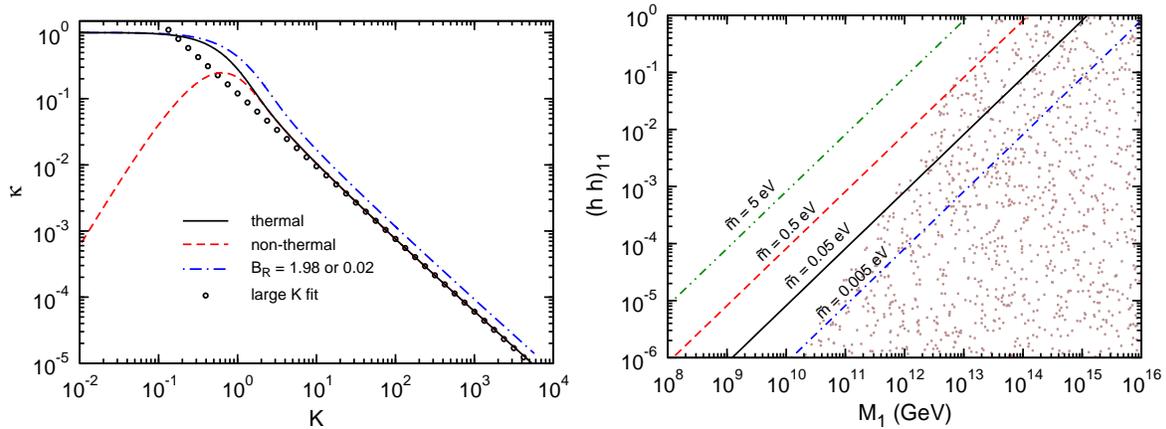

\begin{center}
\includegraphics[scale=0.7]{kappa.eps}
\includegraphics[scale=0.7]{baryo.eps}
\end{center}
\caption{\it (Left panel) Leptogenesis efficiency, $\kappa$ defined in (\ref{kappa}),
  versus K, for thermal and non-thermal initial abundance of $S$ $(\bar{S})$
  and for various $B_R$. (Right panel) Area in the $M_1$, ${\bf (h^\dagger h)}_{11}$
  parameter space allowed by successful baryogenesis when ${\bf (\hnu^\dagger
  \hnu)}_{11} = {\bf (\hp \hp^\dagger)}_{11}$ and $\sigma = v = 175$~GeV.}
\label{fig:kappa}
\end{figure}

After solving the BEs we can still parameterise the baryon asymmetry using the
efficiency $\kappa$ defined in equation (\ref{kappa}). $\kappa$ will depend on
the values of $M_1$, $({\bf \hnu^\dagger \hnu})_{11}$ and $({\bf \hp
  \hp^\dagger})_{11}$. Fig.~\ref{fig:kappa} shows the dependence of $\kappa$
on $K$, for various initial conditions and various values of $B_R$. In cases
with no initial abundance of $S_1$ we see that the maximal efficiency
$\kappa\sim 1$ is indeed reached at $K\sim 1$.

With either an initially thermal abundance of $S_1$, or no initial $S_1$, the
behaviour of $\kappa$ for large $K\gg1$ is the same, and for
$K\stackrel{>}{_\sim} 20$ is well fitted by the power-law
\begin{equation}
\kappa \ \simeq\ \frac{0.12}{K^{1.1}}\ =\ 6.4 \times 10^{-17}\,
\Bigg( \frac{\sigma}{\widetilde{m}} \Bigg)^{1.1}\,.
\label{fit}
\end{equation}
Although much larger CP-asymmetries are required to produced the observed
baryon asymmetry, this large $K$, or `strong wash-out' regime clearly has the
advantage of being insensitive to initial conditions. Fig.~\ref{fig:kappa}
also shows the behaviour of $\kappa$ for differing $B_R$ (or effectively the
ratio $\hnu:\hp$ for small $\delta_R$). $B_R = 1.98$ or $B_R = 0.02$
corresponds to a factor of 10 difference between $\hnu$ and $\hp$. We see that
as $B_R$ departs from 1 the efficiency for large $K$ increases slightly. This
effect is due to the less efficient wash-out via inverse-decays in these
cases, as can be seen in the BEs (\ref{BEs}) where the second term in the
equation for $\eta_{\Delta L}$ responsible for the wash-out of the asymmetry
via inverse decays is also dependent on $B_R$.

On the right hand side of Fig. 3 we show the area in the $M_1$, ${\bf
  (h^\dagger h)}_{11}$ parameter space which is allowed by successful
baryogenesis when ${\bf (\hnu^\dagger \hnu)}_{11} = {\bf (\hp
  \hp^\dagger)}_{11}$ and $\sigma = v = 175$~GeV. The points plotted
correspond to numerical solutions of the BEs with the CP-asymmetry set to the
maximum allowed by the bound (\ref{CPbound}), where the final lepton asymmetry
would result in a baryon asymmetry equal to or exceeding the observed one. On
the plot we also superimpose $\widetilde{m}$ iso-contours.

If we take the most representative natural scenario, namely ${\bf
  (\hnu^\dagger \hnu)}_{11} = {\bf (\hp \hp^\dagger)}_{11} \simeq 1$ and the
reasonable assumption that $\widetilde{m} = 0.05$ eV for hierarchical light
neutrinos we can use the large $K$ fit to the efficiency (\ref{fit}) and the
bound on the CP-asymmetry (\ref{CPbound}) in conjunction with the value of the
observed baryon asymmetry to set an approximate lower limit on $\sigma$.  We
find that unless $\sigma \stackrel{>}{_\sim} 0.1$~GeV the $L_{\nu_R}$
asymmetry produced is insufficient to explain the observed baryon asymmetry.
Notice that this bound depends on several assumptions, in particular that the
heavy $S_i$ are hierarchical in mass. Furthermore, the requirement that the
universe reheats enough to thermally produce the $S_1$ at the end of inflation
leads to an upper bound on $M_1$ of the order of $T_{RH}$, the reheating
temperature. This leads to the approximate upper bound $\sigma
\stackrel{<}{_\sim} 2$~TeV~$(T_{RH}/10^{16}$~GeV$)$.

In summary the scale of the spontaneous symmetry breaking of $U(1)_{\rm D}$ is
bounded,
\begin{eqnarray}
0.1~{\rm GeV} \ \lsim \ \sigma  \  \lsim  \ 2~{\rm
  TeV}\,\Bigg(\frac{T_{RH}}{10^{16}\,{\rm GeV}}\Bigg)\;,
\label{BAU}
\end{eqnarray}
and we can therefore conclude that {\it an electroweak scale $\sigma$ is both
  natural and compatible with successful Dirac leptogenesis.}


\subsection{\it  The Higgs sector }

\subsubsection{\it The potential}

The complete potential of the neutral scalar fields under consideration reads,
\begin{eqnarray}
V(H,\Phi) \ = \ \mu_H^2 H^\dagger H + \mu_{\Phi}^2 \Phi^\dagger \Phi + 
\lambda_H (H^\dagger H)^2 + \lambda_\Phi ( \Phi^\dagger \Phi)^2  - \eta 
H^\dagger H \Phi^\dagger \Phi \;,
\label{potential}
\end{eqnarray}
where all the parameters are real. For the following, we denote $H\equiv H^{0}$.
 Notice that linear or trilinear terms do
not appear thanks to the phantom $U(1)_D$ symmetry.  After $U(1)_D$ is
spontaneously broken, the field $\Phi$ develops a vev $\sigma$, which through
the link $\eta$-term in (\ref{potential}), forces the Higgs field $H$ to also
develop a vev $v$, triggering the ``observed" electroweak $SU(2)_L \times
U(1)_Y$ symmetry breaking\footnote{Notice that the limit $\mu_H \to 0$ is
  attainable and causes no problem for electroweak symmetry breaking.
  However, we cannot justify a possible absence of the $\mu_H$-term from
  (\ref{potential}) by using symmetry or other arguments.}.  Expanding the
fields around the minimum we obtain,
\begin{eqnarray}
H  \ = \   v + \frac{1}{\sqrt{2}} (h + i G)  \quad \;, \quad
\Phi  \ = \ \sigma + \frac{1}{\sqrt{2}} (\phi + i J)  \;.
\label{hf}
\end{eqnarray}
While the Goldstone boson $G$ is eaten by the gauge bosons, the same is not
true for the remaining massless Goldstone boson $J$.  Furthermore, the fields
$h$ and $\phi$, under the influence of the $\eta$-term, mix and become two
physical massive Higgs fields $H_i, i=1,2$ with masses $m_{H2} > m_{H1}$,
\begin{eqnarray}
\biggl ( \begin{array}{c} H_1 \\ H_2 \end{array} \biggr ) \ = \ O \:   
\biggl ( \begin{array}{c} h \\ \phi  \end{array} \biggr ) \qquad {\rm with} \qquad 
O = \biggl ( \begin{array}{cc} \cos\theta & \sin\theta \\ -\sin\theta & \cos\theta \end{array} \biggr ) \;,
\label{rot}
\end{eqnarray}
and mixing angle 
\begin{eqnarray}
\tan2\theta \ &=& \ \frac{\eta v \sigma}{\lambda_\Phi \sigma^2 - \lambda_H v^2} \;.
\end{eqnarray}
The limits $v\gg\sigma$ and $v\ll\sigma$ lead to the usual SM scenario with an
isolated hidden sector.  However, bear in mind that these limits require an
un-naturally small $\eta$ and will not be simultaneously compatible with
neutrino masses and baryogenesis (\ref{BAU}) as advocated in the previous
section. The case $\eta \ll 1$, with phenomenology resembling that of the SM,
 looks like an unnatural corner of the parameter space from the
perspective of the Minimal Phantom SM. However,  the exact limit $\eta=0$ is preserved  
under radiative corrections.
The most interesting scenario is the most ``natural'' one when we require
small Dirac neutrino masses and leptogenesis: $\tan\theta \sim 1$.  If the
phantom sector is responsible for electroweak symmetry breaking then the
natural choice of parameters, when taking into account the positivity
constraint $\lambda_H \lambda_\Phi \ > \ \eta^2/4$, is
\begin{eqnarray}
\lambda_H \sim \lambda_\Phi \sim \eta\sim 1\quad , \quad 
 \tan\theta \sim 1 \quad , \quad \tan\beta \equiv v/\sigma \sim 1  \;.
 \label{nat}
 \end{eqnarray}
 Under these conditions, succesful EW symmetry breaking happens only when 
 $\mu_{\Phi}^2 < 0$~\footnote{At this stage this has to be put in by hand i.e. 
  we cannot offer a radiative mechanism
 to explain this. Dynamical EW symmetry breaking by fourth generation condensates
 has been discussed in~\cite{Paschos}.}. 
 This naturalness condition is supported by an enhanced symmetry $\Phi
 \leftrightarrow H$ of the potential (\ref{potential}) which is broken however
 by the Yukawa couplings and the $U(1)_Y$.  The Higgs potential of
 (\ref{potential}) has been studied in~\cite{Joshipura} in the context of
 Majoron models~\cite{Majoron} where $\sigma$ is an arbitrary vev. This study
 was mainly confined to LEP collider signatures. It is therefore interesting
 to update the phenomenology of this Higgs sector after the LEP era and in
 light of the forthcoming LHC experiments and the condition (\ref{nat}).
 
 From (\ref{rot}) we see that $h=O_{i1} H_i$ and therefore the couplings of
 Higgs bosons $H_i$ to fermions and gauge bosons will be reduced by a factor
 $O_{i1}$ relative to their corresponding SM ones.  It is almost obvious from
 (\ref{potential},\ref{hf},\ref{rot}) that $H_i$ will couple to the
 ``invisible" massless Goldstone pair $JJ$.  The situation is completely
 different to the SM where the $H\to bb$ mode dominates for relatively light
 Higgs masses $\lsim 160$ GeV.  Here we find that in this mass range, the
 decay rate $H_i \to JJ$ relative to $H_i\to bb$ reads as,
\begin{eqnarray}
\frac{\Gamma(H_1 \to JJ)}{\Gamma(H_1 \to bb)} \ &=& \  \frac{1}{48} \biggl (\frac{m_{H1}}{m_b} \biggr )^2
\tan^2\beta \: \tan^2\theta \;,  \label{br1} \\  
\frac{\Gamma(H_2 \to JJ)}{\Gamma(H_2 \to bb)} \ &=& \  \frac{1}{48} \biggl (\frac{m_{H2}}{m_b} \biggr )^2
\tan^2\beta \: \cot^2\theta \;. \label{br2} 
 \end{eqnarray}

\begin{figure}[t] 
   \centering
   \includegraphics[]{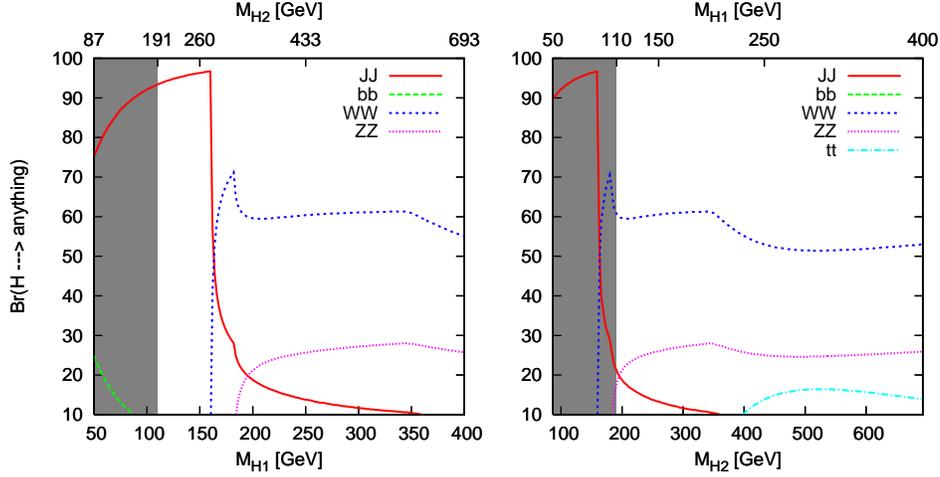} 
   \caption{\em Dominant branching ratios of the two Higgs bosons decaying to invisible $JJ$,  and to SM
     particles $bb$, $WW$, $ZZ$ and $tt$ as a function of the Higgs boson mass
     $H_1$ (left panel) and $H_2$ (right panel) and $\theta = \pi/4$. In every
     case the mass of the partner Higgs boson is displayed in the upper
     horizontal axis for comparison.  The shaded area is excluded by LEP.}
   \label{hfig}
\end{figure}   
Therefore, from (\ref{nat},\ref{br1},\ref{br2}) we see that a light Higgs
boson will decay dominantly to invisible $JJ$ as long as it is heavier than
$60$ GeV.  {\it The existence of the massless Goldstone boson $J$
  diminishes the Higgs boson decay into a $b\bar{b}$ pair.} LHC experimenters
should be aware of this situation which arises in a very simple and natural
extension of the SM! On the other hand, the Higgs decays to SM-vector bosons
and fermions respectively, as
\begin{eqnarray}
\Gamma(H_i \to VV) &=&  \Gamma(H_i \to VV) |_{\rm SM} \times O_{i1}^2 \: \;,
\label{HVV}\\
\Gamma(H_i \to f f) &=&  \Gamma(H_i \to ff) |_{\rm SM}  \times O_{i1}^2 \: \;,
\label{Hff}
\end{eqnarray}
with $V=Z$ or $W$. Analogous formulae are valid for the cross sections
$\sigma(e^+e^- \to V^* \to V H_i)$ and $\sigma(pp \to V^* \to VH_i)$. In the
year 2001, the LEP collaboration presented bounds on an ``invisible" Higgs
boson mass~\cite{LEP}.  Following this analysis, certain Higgs boson mass
values are excluded as a function of the parameter
\begin{eqnarray}
\xi_i^2 \ \equiv \ \frac{\sigma(e^+e^- \to HZ)}{\sigma(e^+e^- \to HZ)|_{\rm SM}} \times {\rm Br}(H \to
{\rm ~invisible} )  =  O_{i1}^2 \times {\rm Br}(H \to
{\rm ~invisible} )\;.
\label{xiparam}
\end{eqnarray}
For $\xi^2=1$, LEP excludes Higgs boson masses up to its kinematical limit,
$m_H \le 114.4$ GeV.  This bound changes only slightly in our case.  To make
the above discussion more concrete we focus on a representative example with
the natural choice of parameters given in (\ref{nat}), i.e.
$\theta=\beta=\pi/4$ and all couplings equal to one. In this case the
production cross sections and branching ratios to gauge bosons, given by
(\ref{HVV}), amount to half of the SM predictions. Thus $\xi_i^2 \sim 1/2$ for
$ {\rm Br}(H \to JJ) \simeq 100$\% which is the case for Higgs masses in the
region $70 \lsim m_{Hi} \lsim 160$ GeV.  This is nicely summarised in
Fig.~\ref{hfig}.  where we plot the branching ratios for both Higgs bosons,
$H_1$ and $H_2$, decaying into invisible Goldstone bosons $J$ and to other
SM-like particles.  In general there are two more ``invisible" decay ``leaks"
not depicted in Fig.~\ref{hfig}: the first is $H_2 \to H_1 H_1$ which is
kinematically forbidden for our choice of $\theta = \pi/4$ and the other is
$H_i \to \nu \nu$ which is proportional to the neutrino masses and is
therefore negligible.  It turns out that for $m_{Hi} \lsim 160$~GeV the LEP
parameter $\xi^2 \gsim 0.4$ and therefore LEP~\cite{LEP} excludes a light
invisible Higgs boson with a mass $m_{H1} \lsim 110$~GeV. This also sets a
lower bound on the partner Higgs boson mass, $m_{H2} \gsim 191$~GeV, which is
now forced to decay only to visible particles $WW, ZZ, tt$. A search for the
latter would follow the SM-type plan, looking for $qqH_2 \to qqWW^{(*)}$ or
$gg\to H_2 \to WW^{(*)}, ZZ^{(*)}$, modes at the Tevatron and LHC.  Regarding
these channels, the only difference here is that production cross sections and
decays are reduced by half.

It is apparent from Fig.~\ref{hfig}, that there is a mass region
\begin{eqnarray}
110 \lsim m_{H1} \lsim  160 ~~{\rm GeV} \;, \label{reg}
\end{eqnarray}
where $H_1$ decays to invisible Goldstone bosons with ${\rm Br}(H_1 \to JJ) >
90\%$.  The question is how can we identify this invisible Higgs boson at the
LHC? This question has been studied extensively in the
literature~\cite{Kane,Stefano,Han}.  The purely invisible Higgs $H_1$ can be
searched for at the LHC through the $Z+H_1$ and/or the W-boson fusion
channels.  Our analysis closely follows the results of \cite{Han} for the
$Z(\to l^+l^-)+H_{\rm inv}$ production mode, where we multiply their
$S/\sqrt{B}$ by a factor of 1/2 because of (\ref{HVV}). We find that for an
LHC integrated luminosity of $30\;{\rm fb}^{-1}$ the signal significance for
the invisible $H_1$ with a mass of 120 (140) [160] GeV is $4.9\sigma (3.6
\sigma) [2.7\sigma]$ respectively.  Although these results refer to the case
where $\theta=\pi/4$ the situation is rather generic in the region of
(\ref{nat}). Although the above analysis for the Higgs boson decay to
invisible is very important to identify the nature of the phantom sector one
may also identify the light Higgs boson when $m_{H1}\lsim 140$ GeV through the
conventional $H_1 \to \gamma\gamma$.  We have not made a detailed study for
this mode and we believe that it is worth further investigation. For
$m_{H1}>160$~GeV, $H_1$ decays mainly to $WW$ and $ZZ$ since the Br($H\to JJ)
\simeq 1/13$ is suppressed. Notice however that $m_{H1} \gsim 200$ GeV is
rather disfavoured by the EW data as we will see shortly.

In conclusion, the phantom sector of (\ref{potential}) allows for Higgs decays
into invisible Goldstone scalars and to visible gauge bosons. {\it A situation
  may arise where LHC experimenters could detect $H_1$ with $m_{H1} \lsim
  120$~{\rm GeV} through the invisible $Z(\to l^+l^-) + H_1$ mode and $H_2$ with
  $m_{H_2} \simeq 200$~{\rm GeV} through its production and decays in association
  with gauge bosons.}

\subsubsection{\it The $\rho$-parameter and other observables:}

In a model with only Higgs doublets and singlets, the tree level value for the
electroweak parameter, $\rho \equiv m_W^2/M_Z^2 \cos^2\theta_W$, is
automatically equal to one without further adjustment of the parameters of the
theory. The correction to the parameter $\rho$, denoted by $\Delta \rho$,
appears at one loop level.  For the model at hand, the phantom singlet $\Phi$
will affect gauge boson loops through its $\eta$-mixing term with the
observable Higgs field $H$. Then it is straightforward to calculate the Higgs
contribution to $\Delta \rho$~\cite{Ramond}. It reads,
\begin{eqnarray}
\Delta \rho^{H} \ = \ \frac{3 G_F}{8 \sqrt{2} \pi^2} \sum_{i=1}^2 O_{i1}^2 \biggl [
m_W^2 \ln \frac{m_{Hi}^2}{m_W^2} - m_Z^2 \ln \frac{m_{Hi}^2}{m_W^2} \biggr ]
\;,
\label{drho}
\end{eqnarray}
where $O$ is the orthogonal matrix in (\ref{rot}). We can establish  a useful
 connection between this formula and the SM one. Note 
 that from the similarity condition of the rotation matrix, 
$O^T m^2 O = diag(m_{H1}^2, m_{H2}^2)$, with  $m$ being the 
$2\times 2$ Higgs mass matrix, we read the following identity,
\begin{eqnarray}
\sum_{i=1,2} m_{Hi}^2 O_{i1}^2 \  = \ 4 \lambda_H v^2 \  \equiv \ m_H^2 
\;,
\label{sim}
\end{eqnarray}
where $m_H$ is the SM Higgs boson mass expression.  It is easy now to simplify 
our expression  for $\Delta \rho$, by Taylor expanding (\ref{drho})  around $m_H^2$,
\begin{eqnarray}
\sum_{i=1}^2 O_{i1}^2 f(m_{Hi}^2) = \sum_{i=1}^2 O_{i1}^2  \biggl [f(m_H^2) \ + \ 
(m_{Hi}^2 - m_H^2) f'(m_H^2) \ + \ ... \biggr ] \;,
\label{taylor}
\end{eqnarray}
where $f(x)$ is a continuous function and $f'(x)$ denotes its
derivative with respect to $m_{Hi}^2$.
 Using  (\ref{sim}) and the orthogonality condition $O^T O = 1$, 
 the second term in (\ref{taylor}) vanishes identically, leading to
\begin{eqnarray}
\Delta \rho^{H} \ = \ \frac{3 G_F}{8 \sqrt{2} \pi^2}  \biggl [
m_W^2 \ln \frac{m_{H}^2}{m_W^2} - m_Z^2 \ln \frac{m_{H}^2}{m_W^2} \biggr ] \;,
\label{drhosm}
\end{eqnarray}
which is just the SM Higgs contribution to $\Delta\rho$.  One arrives at the
same conclusion for the $S,T$ and, $U$ parameters. Assuming that the Higgs
contributions to the non-oblique corrections follow the same pattern, we can
use the electroweak constraint on the SM Higgs boson mass, $m_H<194$ GeV at
95\% C.L~\cite{Erler} in order to set constraints on the Higgs boson masses
and mixing angle of this model. Thus, comparing  (\ref{drho}) with  (\ref{drhosm})
we arrive at  
\begin{eqnarray}
\cos^2 \theta \log (m_{H1}^2) \ + \ \sin^2\theta \log (m_{H2}^2) \ <  \   \log (194^2 ~{\rm GeV}^2)
 \quad {\rm (at ~} 95\% ~{\rm C.L.).}
 \label{ewc}
\end{eqnarray}
In the case of our working example $\theta = \pi/4$, this translates into
$m_{H1} m_{H_2} < 194^2~{\rm GeV}^2$, e.g, 
$m_{H1} \lsim115$ GeV and $m_{H2} \lsim 327$ GeV. These bounds have to be
combined with the LEP bounds on the Higgs masses derived in the previous
section.

It is apparent that the inclusion of the phantom singlet field $\Phi$, does not
affect the GIM mechanism~\cite{GIM} which is responsible for the absence of tree level
flavour changing neutral currents (FCNC). One may think of Higgs mediated contributions to
rare B-decays at loop-level.  The most striking one would have been: B(or Y)-meson 
decays to invisible, $B\to JJ$.  Alas, the amplitude for this decay is
proportional to $\sum_i O_{i1}O_{i2}$ which vanishes because the matrix $O$ is
orthogonal. This is a kind of GIM mechanism suppression in the Higgs sector. 
Other Higgs mediated contributions to observables like $B\to \mu^+
\mu^-$ or to the muon anomalous magnetic moment, $g-2$, will follow the SM
prediction thanks to the relation (\ref{taylor}) and the bound on the light
Higgs mass (\ref{reg}).  In conclusion, {\it the minimal singlet phantom
  sector does not change the FCNC predictions for processes existing in the SM.}

\subsection{\it  Cosmological and astrophysical constraints}

Let us finally address the implications for cosmology and astrophysics.  The
presence of the massless Goldstone boson, $J$, has interesting consequences
which need to be analysed.

During the expansion of the Universe, a critical temperature is reached below
which the $U(1)_D$ symmetry is spontaneously broken. As we have already
explained in the previous section, the field $\Phi$ then develops a
non-vanishing vev and its real part, $\phi$, mixes with the real part of the
SM Higgs field, giving rise to two scalar Higgs mass eigenstates, $H_i$.  On
the other hand, $J$ survives as the massless Goldstone boson.  From
(\ref{potential},\ref{hf},\ref{rot}) $H_i$ couples to the Goldstone pair $JJ$
as
\begin{eqnarray}
  -{\cal L}_J&\supset&
  \frac{(\sqrt{2}G_F)^{1/2}}{2}\tan\beta\,
  O_{i2}\,m_{H_i}^2\ 
  H_i\,J J\ .
  \label{hjj}
\end{eqnarray}
The Goldstone bosons are then
kept in equilibrium via reactions of the sort
$JJ\leftrightarrow f\bar f$,
mediated by $H_i$. However, 
since the amplitudes for these processes are
suppressed 
due to a GIM-like mechanism which stems from the
orthogonality condition, $\Sigma_i O_{i1}O_{i2}=0$, of the matrix $O$, 
$J$ falls out of equilibrium before the QCD phase transition
and remains as an extra relativistic
species thereafter.

The presence of relativistic particles, apart from neutrinos, is strongly
constrained by Big Bang Nucleosynthesis (BBN), since they alter the predicted
abundances for the light elements.  Namely, additional relativistic particles
would increase the expansion rate of the Universe, leading to a larger
neutron-to-proton ratio and therefore a larger $^4$He abundance.  The allowed
number of extra relativistic degrees of freedom is usually parameterised by
the effective number of neutrino species, $N_{eff}=3+\Delta N_{\nu}$.
Observations of the primordial $^4$He abundance, combined with the CMB
determination of the baryon-to-photon ratio yield $N_\nu=3.24\pm 1.2$ at 90\%
CL \cite{cyburt,fieldssarkar}, and 
a similar upper bound on $N_\nu$ can be derived from 
analysis of the CMB and large scale structure 
\cite{hann}.
Notice that this does not pose a problem for
$J$s. Since they decoupled at a temperature above the QCD phase
transition, 
when $g_*\gsim60$,
their temperature at BBN, $T_J$, is smaller than that of neutrinos and
photons, $T$. Namely $(T_J/T)^4\lsim(10.75/60)^{4/3}$.  This is equivalent to
an increase in the number of neutrino species of just $\Delta
N_\nu=4/7\,(T_J/T)^{4}\lsim 0.06$, well in agreement with the
above-mentioned constraints.

On the other hand, $J$ also induces the decay of heavy neutrinos into the
lightest one, $\nu_H\to J\,\nu_L$. Their interaction is described by the
following effective Lagrangian,
\begin{eqnarray}
  {\cal L}_{J\nu\nu}  & \supset & \frac{{\bf \hat{m}_\nu}}{\sigma} \,
  i \: {\bf \bar{\nu} } \:  
  \gamma_5  \: {\bf \nu} \: J \;,
  \label{lagrphi}
\end{eqnarray}
with ${\bf \hat{m}_\nu}$ a diagonal matrix which contains the physical
neutrino mass eigenstates.  Since the decay does not include any photon in the
final state, some potential cosmological problems associated with radiative
neutrino decays (e.g., contributions to the diffuse photon background and
distortions to the cosmic microwave background black body spectrum) are
avoided.  However, limits on the non-radiative decay of neutrinos (usually
expressed as upper bounds on the $J\nu\bar\nu$ coupling) can be derived from
solar neutrino observations ($g_{J\nu\bar\nu}^2\lsim10^{-5}$)
\cite{beacombellband}, meson decays ($g_{J\nu\bar\nu}^2\lsim10^{-4}$)
\cite{meson}, as well as from preventing overcooling in supernovae (which
exclude a range around $g_{J\nu\bar\nu}^2\lsim10^{-10}$, although the bounds
are model dependent) \cite{overcooling}.

As we can read from (\ref{lagrphi}), in the present model
$g_{J\nu\bar\nu}={m_\nu}/{\sigma}$. For natural (electroweak scale) values of
$\sigma$, and with $m_\nu\lsim1$ eV we obtain
$g_{J\nu\bar\nu}^2\lsim10^{-22}$, thus fulfilling all the aforementioned constraints.

Alternatively, a calculation of the heaviest `light' neutrino lifetime, in the
case of hierarchical neutrino masses, yields
\begin{equation}
\frac{\tau_\nu}{m_\nu}=\frac{16\pi}{g_{J\nu\bar\nu}^2m_\nu^2}\sim{\cal
  O}(10^{13})\ {\rm s/eV},
\end{equation}
where $\sigma\sim100$ GeV and $m_\nu\sim0.05$ eV, has been used. A similar
value is obtained in the opposite limit, when neutrino masses are
quasi-degenerate.  The heaviest `light' neutrino is therefore extremely
long-lived and escapes all the constraints on neutrino decays.

Finally, notice that the Goldstone boson $J$ couples very weakly to electrons,
 through one loop  diagrams,
with strength $g_{eJ} \simeq G_F m_\nu^2 m_e/\sigma$, and it does not affect
the evolution of stars~\cite{dearborn} if $g_{eJ} \lsim 10^{-12}$ which
implies that $\sigma \gsim 5\times 10^{-17}$ GeV.
Also, the emission of a Goldstone pair
mediated by (virtual) Higgses 
is negligible in our
model, once more due to the orthogonality condition of (\ref{rot}),
and the strong constraints on the Higgs couplings, obtained
from studies on star evolution
\cite{bertolini}, are trivially fulfilled.

\subsection{\it Conclusions}

We have proposed the minimal, lepton number conserving, SU(3)$_{\rm
  c}\times$SU(2)$_L\times$U(1)$_Y$ gauge-singlet, or phantom extension to the
Standard Model leading to naturally small Dirac masses for the neutrinos and
baryogenesis through Dirac leptogenesis. The extension is natural in the sense
that all couplings are either of ${\cal O}(1)$ or strictly forbidden.

Spontaneous breaking of a global, phantom sector U(1)$_{D}$ symmetry will
trigger electroweak symmetry breaking. The scale of this phantom sector
symmetry breaking is constrained to be around the electroweak scale by the
simultaneous requirement of successful Dirac leptogenesis and small light
neutrino masses. In this model, small Dirac neutrino masses arise through a
mechanism very similar to the standard Majorana see-saw. The model can also be
viewed as a very simple Froggatt-Nielsen scenario of the sort usually invoked
to generate large hierarchies in the quark masses.

Baryogenesis through Dirac leptogenesis occurs naturally in this model since
the small {\it effective} Yukawa couplings of the left and right-handed
neutrinos prevent the left and right neutrino asymmetries from equilibrating
once they are created.  The initial neutrino asymmetry is created via the out
of thermal equilibrium decays of the heavy Dirac particles $S_i$ and
$\bar{S}_i$, in analogy with conventional Majorana leptogenesis.

A Davidson-Ibarra-like bound on the CP-asymmetry in the $S_i$, $\bar{S}_i$
decays exists when their masses are hierarchical. This bound, in conjunction
with information on the efficiency of leptogenesis extracted from the solution
of the Boltzmann equations allows us to place a lower bound on the vev of the
phantom sector, SM gauge-singlet $\Phi$, such that the asymmetry created in
Dirac leptogenesis is enough to explain the observed baryon asymmetry of the
Universe. Assuming a hierarchical light neutrino spectrum and a hierarchical
$S_i$ mass spectrum, we find that
\begin{equation}
\sigma \gsim 0.1\;{\rm GeV}\,.
\end{equation}
Making the further assumption that leptogenesis must proceed after the thermal
production of the $S_1$ following a period of inflation leads us to an
approximate upper bound on $M_1$ and therefore $\sigma$
\begin{equation}
\sigma \lsim 2\;{\rm TeV}\,\Bigg(\frac{T_{RH}}{10^{16}\,{\rm GeV}}\Bigg).
\end{equation}
Thus we find that an electroweak scale $\sigma$ is simultaneously compatible
with both light neutrino data and successful Dirac leptogenesis. Significantly,
an electroweak scale $\sigma$ is also required by our naturalness criterion,
since the mixing of the $\Phi$ and the SM Higgs is expected to be maximal.

The addition of the phantom sector scalar $\Phi$, which mixes with the SM
Higgs, introduces an additional massive Higgs boson. After the breaking of the
global U(1)$_D$, we are also left with a massless Goldstone boson, $J$.  This
Goldstone boson couples to the two physical Higgs bosons introducing an
additional, invisible decay mode for the Higgs $H_i \to JJ$.  This decay mode
suppresses the branching ratio of Higgs to $bb$ and instead both Higgs bosons
decay dominantly to invisible $JJ$ and/or to vector boson pairs.
We discuss in detail a natural scenario with a representative mixing angle
($\theta=\pi/4$) and estimate that with 30~fb$^{-1}$ of integrated luminosity
the LHC could find the invisible Higgs with a mass 120~GeV, at a significance
of $4.9\,\sigma$.  In addition, at the same time a significantly heavier,
partner Higgs could be found with a mass 200~GeV through its vector boson
decays. Interestingly, electroweak constraints suggest an upper limit of $\sim
250$~GeV for the mass of the Higgs bosons.

The model passes relevant FCNC constraints thanks to a GIM-like mechanism.
Cosmological bounds on the number of relativistic species at BBN are also
fulfilled, due to the Goldstone boson decoupling before the QCD phase
transition.  Finally, astrophysical constraints on the Goldstone couplings
from neutrino decays and stellar evolution are trivially satisfied.

\vspace*{0.5cm}
\noindent {\large \bf Acknowledgements}

D.C thanks K.~Jedamzik for very helpful discussions. A.D would like to thank
the Nuffield Foundation for financial support. D.C. and T.U. thank PPARC for
financial support.


\end{document}